\documentclass[11pt]{article}

\usepackage[a4paper,margin=1in]{geometry}
\usepackage{amsmath,amssymb,amsfonts}
\usepackage{mathtools}
\usepackage{physics}
\usepackage{hyperref}
\usepackage{enumitem}
\usepackage{graphicx}

\newcommand{\HU}{\mathcal{H}_{\mathrm U}}
\newcommand{\HW}{\mathcal{H}_{W}}
\newcommand{\HMi}{\mathcal{H}_{M_i}}
\newcommand{\HR}{\mathcal{H}_{R}}
\newcommand{\Id}{\mathbb{I}}
\newcommand{\HM}{\bigotimes_{i=1}^n \mathcal{H}_{M_i}}

\title{What It Is Like To Be A Quantum Computer:
A Closed Quantum Global Workspace - Conscious Access and Integration as Correlation Dynamics in Hilbert Space}
\author{
Libby Heaney\\
\small Independent artist, PhD Quantum Information Science, \\
\small London, United Kingdom\\
\small \texttt{studio@libbyheaney.co.uk}
}
\date{\today}

\begin{document}
\maketitle

\begin{abstract}
Motivated by speculative artistic inquiry into quantum-only forms of experience in relation to representation and embodiment, this paper develops a translation of the Global Neuronal Workspace (GNW) theory for closed quantum systems. The work treats classical GNW as an architectural reference and asks how conscious access and memory change under unitary dynamics.

We reference a quantum Hopfield-style Hamiltonian that mirrors key organisational features of GNW,
such as a distributed workspace and tunable large-scale integration.
Within this framework, global availability of information is realised through coherence-preserving correlation distributions and memory is a persistent internal correlation structure.
Likewise, ignition becomes a transition in global entanglement as opposed to convergence towards a stable attractor. Reductions of the Hopfield-style Hamiltonian lead to a tractable toy model and GHZ-type entanglement, explicitly showing how relative phase is globally accessible through joint internal operations yet absent from all local subsystems.

This closed quantum system approach clarifies which aspects of workspace theories are contingent on classical assumptions. Our model highlights a non-representational regime in which access and memory are entirely relational rather than copies or records.
Beyond its conceptual contribution, the paper proposes a mode of art-science collaboration where speculative artistic inquiry is pursued through theoretical models rather than illustration or data.
\end{abstract}

\tableofcontents

\section{Introduction}

\subsection{Motivation through artistic inquiry and quantum theory}

This paper was motivated by the author's artistic research exploring quantum aesthetics in relation to representation and embodiment. 
Much of the history of representational art has developed in dialogue with the classical world of objects with recordable positions and trajectories.
However, representational strategies that capture images of objects described by classical physics struggle to express quantum phenomena such as superposition and entanglement, which typically exist in high dimensional Hilbert spaces \cite{nielsen_chuang}. 
The author's art practice grapples with this, resulting in works like {\it{Ent-}} (2022) \cite{las_ent}, where recognisable digital forms are animated by coherence data from various quantum algorithms and tend towards abstraction rather than depiction.

While artistic abstraction offers one approach to the slippery problem of representing quantum phenomena, embodied experience may offer alternative routes. 
Rather than asking how quantum systemsit {\it look}, how might one might {\it feel} quantum superposition and entanglement through embodiment?
In other words, what might it feel like to experience a quantum-only ontology or to exist inside, for example, an error-corrected quantum computer? 
Conversely, what might it mean to experience the internal organisation of a closed quantum system from its own perspective?

Questions like this lead us to longstanding philosophical discussions of subjective experience, often framed as the problem of ``what it is like'' to be a system \cite{nagel1974}.  In its most general expression, we are asking ``{\it{What is it like to be a closed quantum system?}}''. In other words, what is it like to be an error-corrected quantum computer?
In contemporary philosophy of mind, this issue is closely related to the notorious hard problem of consciousness: explaining how physical processes give rise to subjective experience \cite{chalmers1995}. 
Contemplating these questions led the author to consider whether quantum-only forms of consciousness might be conceivable and how artistic inquiry and quantum models might contribute to this. 

The author is a practicing, internationally exhibited artist with formal training in quantum information science (MSci Physics from Imperial College London; PhD from University of Leeds and post-doctoral fellowships at the University of Oxford and the Centre for Quantum Technologies at the National University of Singapore as well as an MA Art and Science from Central St. Martins, London). This dual background, together with the sustained development of visual work using IBM’s publicly available quantum processors, beginning in 2019, already led to a new affective understandings of quantum-driven artworks. 
However, the author's interest in quantum-only ontologies meant that the theoretical literature grounding her art  was entirely absent (see appendix \ref{app} for further background). This paper bridges this gap. It primarily adopts a technical approach, with aspects of the subsequent discussion drawing on insights from visual art. A sister paper that approaches these questions through the author's artistic practice will be published elsewhere.

Thus, on the one hand, the objective, reductionist methods of the natural sciences are unlikely to fully capture subjective experience \cite{nagel1974}. In human consciousness, the transition from the underlying structures \cite{schiff2010} and neural correlates \cite{casali2013} to felt experience, i.e. {\it``what it is like to be''} something, remains unresolved.

Consciousness in artificial intelligence research has recently become an increasingly rigorous and active field of investigation \cite{butlin2023consciousness}. However, almost all such work assumes classical computational architectures. 
While a variety of ``quantum consciousness'' hypotheses have been proposed, they typically involve quantum processes in largely classical biological systems, \cite{hameroff_penrose1996, gassab2025quantum}. How architectural principles of consciousness might manifest in closed quantum systems is an open question.

On the other hand, art in its many forms cannot make specific, testible empirical claims about underlying mechanisms.
Art, however, does engage with the feelings and sensations of subjective experience: what its like to be something. 
Artistic practices often explore perception, embodiment, affect and forms of being that resist purely objective description. 
For this reason, artistic inquiry plays a crucial role not only in motivating questions about consciousness that extend beyond the explanatory scope of current scientific models, but also in constructing experiential encounters that provide approximations to phenomena that resist direct representations. Together they offer complementary ways of approaching questions about experience that neither domain may fully resolve alone.

To begin, this paper takes key features of the Global Neuronal Workspace (GNW) model of conscious access and asks what remains, what changes and what disappears when its core features are translated into quantum-only, never-classical systems \cite{dehaene2011experimental,mashour2018conscious}. 

The introduction concludes with a summary of the main scientific contributions of the paper and an overview of its structure. 
Subsequent sections develop the quantum reformulation of the Global Neuronal Workspace framework, present the resulting closed quantum `GNW' toy model and discuss the implications of closed quantum system conscious access and memory.

\subsection{Overview and key results}

The paper develops a constraint-led reformulation of the Global Neuronal Workspace into a fully unitary, Hilbert-space-native architecture. To be clear at the outset, this paper is not trying to claim the existence of consciousness in, eg error corrected quantum computers, rather it is inspired by artistic inquiry around quantum-only ontologies and speculations on what quantum consciousness might look and feel like if it did exist. 

The core contribution of this paper is not the introduction of new models per se, as related models have already been explored within the quantum machine learning context \cite{hopfield1982neural,rebentrost2018quantum, miller2021qham}. Rather, the novelty lies in reformulating these models in terms of GNW-inspired conscious access and memory. Our claims are as follows:

Firstly, after reviewing core features of classical GNW,
we show that classical broadcast cannot exist in closed quantum systems.
Instead, global availability is realised through coherence-preserving
correlation distributions. Information becomes globally available through entanglement between subsystems, rather than through duplication as in the classical GNW model.

Secondly, we demonstrate that relative phase plays a central role
as a globally accessible invariant.
Phase information is absent from all local subsystems, yet present in global correlations across the entire system, showing how access can be global without being locally representable.

Thirdly, we redefine memory in structural terms.
Without decoherence and classical records, memory cannot be stored as narratable information.
Instead, memory is identified with a persistent internal correlation structure that constrains future dynamics.
Quantum memory is thus intrinsically non-narratable.

Fourthly, we reformulate ignition as a transition in global correlation structure under unitary dynamics.
A toy  model of a transverse-field Ising illustrates how such transitions can arise without measurement or noise \cite{sachdev_qpt}.

These results imply that in a closed quantum model of GNW there are no pictures and no narratives.
What remains are relational invariants (patterns of phase and entanglement) that shape the system’s ongoing evolution.
Scientifically, this clarifies which elements of GNW are contingent on classicality and how they are reformulated in Hilbert space.
Artistically, it delineates a space of possible experience that is structural and relational, rather than representational or narrative.

Finally, we stress this work does not propose a model of biological or human consciousness.
It asks a counterfactual question. What would it mean, in strictly quantum-information-theoretic terms,
for a system to exhibit global access, integration and memory analogous to ``workspace'' architectures studied in cognitive science, if classicality never emerged?

Finally, recent work has argued that decision-making and agency require hybrid quantum–classical regimes with measurement and basis selection \cite{Adlam25}. The present work deliberately excludes such regimes. This is the first time conscious access has been investigated in a closed quantum system.

\section{Classical Global Neuronal Workspace}

\subsection{Background: Classical Global Neuronal Workspace (GNW)}

\begin{figure}
\centering
  \includegraphics[width=0.66\linewidth]{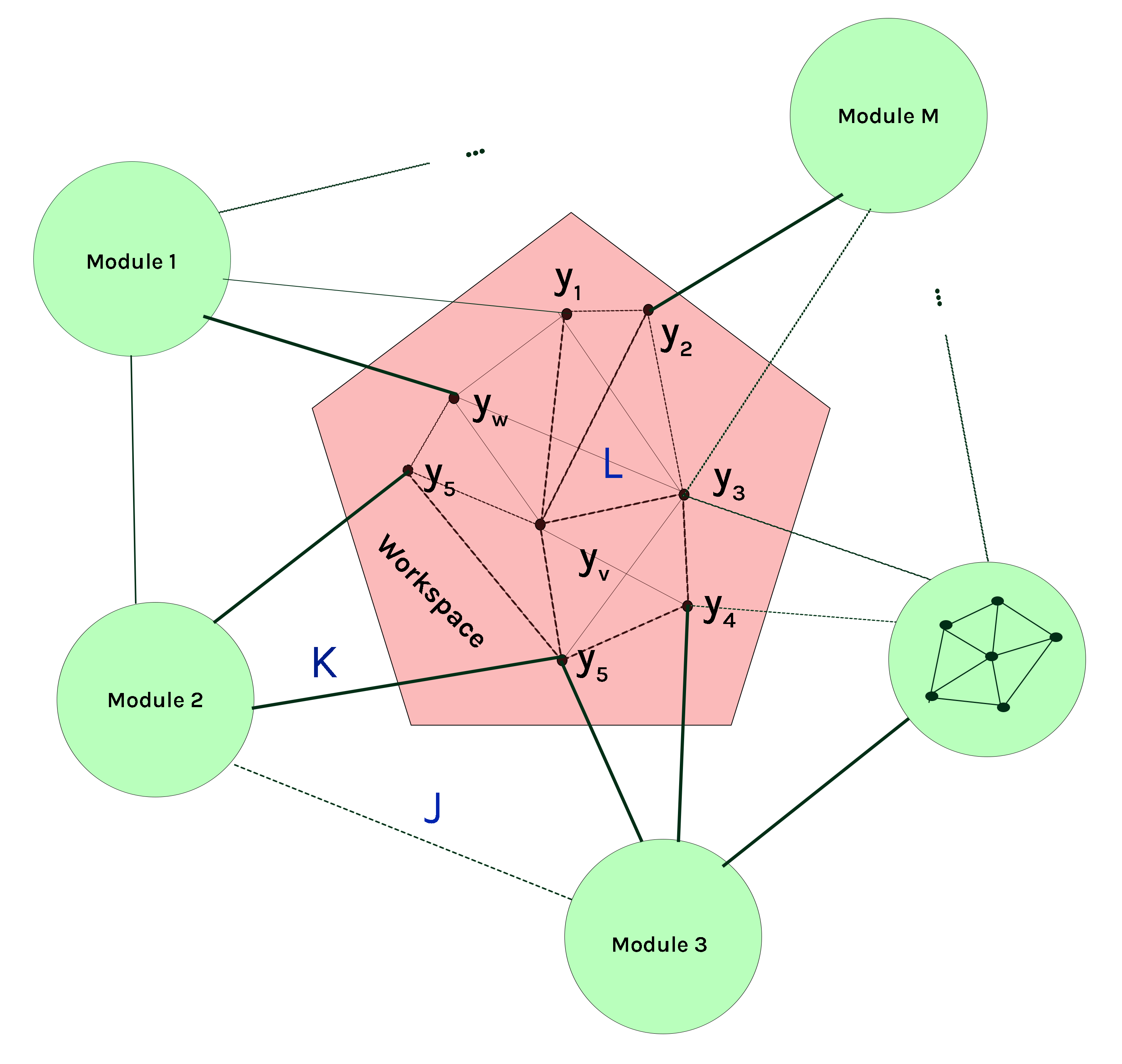}
  \caption{Classical GNW model. Connections of the modules to the nodes of the workspace and weight matrices, L, K and J.}
  \label{fig:GNW}
\end{figure}

The Global Neuronal Workspace (GNW) is a theoretical framework for conscious access
in which typically unconscious information processed by specialised subsystems of neurons called modules becomes globally available and conscious through distributed, recurrent workspace networks \cite{dehaene2011experimental,mashour2018conscious, Tavares2025modeling}, see Fig. \ref{fig:GNW}.
In its classical formulation, GNW is typically modelled using population-level variables that describe the activity of neural assemblies, rather than individual neurons or synapses. 

A module is a specialised neural subsystem that processes a particular type of information largely outside of conscious awareness, for instance facial recognition, language parsing or auditory planning.
 A workspace node is a functional unit within the distributed network that enables information to become globally available and integrated across the brain when information becomes conscious.

Let $x_i(t)$ denote the mean activity of module $i$,
and let $y_\alpha(t)$ denote the activity of  workspace node $\alpha$.
A broad class of GNW-inspired neural models can be written as
\begin{equation}
\label{GNW1}
\tau \dot{x}_i (t) = -x_i(t)  + S\!\left(\sum_j J_{ij} x_j(t)  + \sum_\alpha K_{i\alpha} y_\alpha(t)  + I_i\right),
\end{equation}
and similarly
\begin{equation}
\label{GNW2}
\tau_W \dot{y}_\alpha(t)  = -y_\alpha (t) + S\!\left(\sum_\beta L_{\alpha\beta} y_\beta (t) + \sum_i K_{\alpha i} x_i (t) + I_\alpha\right),
\end{equation}
where $J_{ij}$, $L_{\alpha\beta}$ and $K_{i\alpha}$ encode effective connectivity,
$I_i$ and $I_\alpha$ denote inputs,
$S(\cdot)$ is a nonlinear activation (often sigmoidal),
and $\tau,\tau_W$ are time constants. The connectivities $J, L$ and $K$ are independent of time. The dynamics are entirely encoded in the activity variables $x$ and $y$.

In GNW, global availability arises when recurrent excitation between workspace and modules drives the system
through a nonlinear threshold, producing a transition from brief local activity to a stable, self-sustaining, distributed pattern of coordinated activity that persists beyond the initial stimulus.
This transition, \emph{ignition}, can correspond to a bifurcation from a low-activity fixed point to a stable high-activity attractor \cite{dehaene2011experimental,mashour2018conscious}.

Classical GNW relies on three features:
(i) nonlinear activation functions that produce threshold behaviour,
(ii) dissipation and noise that stabilise attractors,
and (iii) classical state variables that support copying of information.
These assumptions are well suited to biological neural systems,
but will be explicitly reformulated in the closed quantum GNW-like model.

\subsection{Background: Why GNW is attractive}
The structure of GNW:
(i) \emph{specialised processors} (modules),
(ii) \emph{global availability} via long-range recurrent connectivity,
and (iii) \emph{ignition} as a thresholded transition into a self-sustaining, globally broadcast state \cite{dehaene2011experimental,mashour2018conscious}.
give rise to the behavioural and neurophysiological signatures
of access (e.g.\ all-or-none dynamics, late global recurrence, reportability, working-memory maintenance etc).

Furthermore, GNW instantiates a thresholded amplification mechanism
via nonlinear firing-rate functions $S$ embedded in recurrent loops that yield bistability and winner-take-all selection, which are natural for modelling ``one percept at a time'' and abrupt access transitions.

\subsection{Background: Limitations of GNW as a general theory of consciousness in brains}
GNW is not generally presented as a complete theory of consciousness.
Rather, it is a theory of \emph{conscious access}. It shows the conditions under which information becomes globally available for reporting and working memory \cite{dehaene2011experimental,mashour2018conscious}.
Several limitations are widely discussed.

Firstly, GNW focuses on access-related signatures and is comparatively silent on the intrinsic character of felt experience.
Secondly, GNW is historically tied to the ability to report and conduct tasks.
Although ``no-report'' approaches aim to dissociate report from consciousness, the strongest GNW-aligned results involve explicit behavioural report \cite{tsuchiya2015no,block2019consciousness,boly2017are}.

Given these limitations, one might ask why GNW is chosen over other classical models on consciousness as the basis for a quantum reformulation.
Firstly, we are not motivated by discussions of human brains, merely speculating around general features of quantum consciousness compared to classical cases. Therefore, it does not really matter which classical model we choose to reformulate, just that it begins to indicate how key features begin to change.
Secondly, from a systems perspective, GNW offers a well-supported description of how distributed subsystems become functionally integrated at macroscopic scale, which can be meaningfully reformulated in Hilbert space.
GNW's core features are not tied to specific neuronal mechanisms,
but rather to abstract organisational principles, which can be re-expressed under different physical constraints.

An additional motivation is that GNW poses concrete questions about representation and narrative.
These questions are interesting for artistic inquiry, because they indicate the forms of experience and meanings that might be possible in systems whose organisation is fundamentally non-classical.

\subsection{Limitations of GNW in the quantum domain}

The following identify which core assumptions of classical GNW fail under closed unitary dynamics, motivating the specific reformulations in subsequent sections.

\paragraph{ Classical broadcast presumes copying:}
Classical global availability is often discussed as ``broadcast'' of selected content to many subsystems.
But in a quantum setting, arbitrary state copying is forbidden by no-go theorems \cite{wootters_zurek_1982, nielsen_chuang}. In closed quantum systems, global availability cannot be through classical copies.

\paragraph{ Ignition is typically modelled as a dissipative attractor:}
In recurrent-network implementations, ignition corresponds to a nonlinear bifurcation into a stable attractor.
Such convergence relies on dissipation/noise and coarse-grained classical state spaces.
Unlike classical neural systems, finite closed quantum systems evolve unitarily and therefore do not generally converge to attractors.

\paragraph{Reportability and narratable memory are classical phenomena:}
GNW results often speak of report and stable representations.
However, in a closed quantum system evolve unitarily, rather than through outcome selection.
 ``Memory'' cannot be classical records without reintroducing measurement.
Thus a quantum reformulation of GNW must redefine memory.

\section{A quantum Hopfield-style global workspace}

\subsection{Hopfield-style Hamiltonian}
Later in this paper, we highlight toy constructions that evidence key features with minimal algebraic overhead.
But first, we write down a closed-system Hamiltonian mirroring the classical GNW formulation in that it (i) supports distributed, recurrent interactions
and (ii) allows an explicit ``workspace-module'' decomposition with tunable coupling.

A natural candidate is a quantum Hopfield-style model, which generalises the classical Hopfield energy function into an Ising-type Hamiltonian. This has been explored in several forms in quantum information and quantum machine learning contexts
\cite{hopfield1982neural,rebentrost2018quantum, miller2021qham}.

Let $W$ denote a workspace register of $N_W$ two-level systems, or qubits,
and let the modules be partitioned into $n$ groups with sizes $N_{M_i}$.

The Pauli operators $X_j,Y_j,Z_j$ act on qubit $j$ in the computational basis.
All dynamics is generated by a time-independent or time-dependent Hamiltonian $H(t)$
on the closed Hilbert space. 

A Hopfield-style construction begins from an Ising energy function whose low-energy states
encode stored ``patterns''. In a quantum setting, this appears as a Hamiltonian
with diagonal $Z$-$Z$ couplings plus non-commuting terms that preserve coherence.
\\
We write the total Hamiltonian as
\begin{equation}
H(t) \;=\; H_W \;+\; \sum_{i=1}^n H_{M_i} \;+\; H_{WM}(t) \;+\; H_{\mathrm{drive}}(t).
\label{eq:H_total_hopfield_gnw}
\end{equation}
\\
\paragraph{Workspace as a long-range recurrent register}
We take a Hopfield/Ising form on the workspace qubits,
\begin{equation}
H_W \;=\; -\frac{1}{2}\sum_{\alpha\neq\beta}^{N_W} J^{(W)}_{\alpha\beta}\, Z_\alpha Z_\beta
\;-\;\sum_{\alpha=1}^{N_W} b^{(W)}_\alpha\, Z_\alpha.
\label{eq:H_workspace}
\end{equation}
where $J^{(W)}_{\alpha\beta}$ is a dense, and in general long-range, coupling matrix.
In the GNW analogy, this term plays the role of the recurrent workspace connectivity $L$
that supports global integration.
\\
\paragraph{Modules as specialised subsystems}
For each module $i$, we write
\begin{equation}
H_{M_i} \;=\; -\frac{1}{2}\sum_{\substack{a\neq b\\ a,b\in M_i}} J^{(i)}_{ab}\, Z_a Z_b
\;-\;\sum_{a\in M_i} b^{(i)}_a\, Z_a,
\label{eq:H_module_i}
\end{equation}
where $J^{(i)}$ encodes intra-module couplings
and $b^{(i)}_a$ are local bias terms. In the GNW analogy, these terms play a role similar to baseline excitability
or task-dependent tuning.\\

\paragraph{Workspace-module coupling}
In classical GNW models, global availability is mediated by long-range,
bidirectional couplings between workspace nodes and specialised modules,
encoded by the $K_{i\alpha}$ and $K_{\alpha i}$ terms
in the classical neural equations introduced earlier Eqs.~\eqref{GNW1}\eqref{GNW2}.
These couplings allow selected module-level activity to influence workspace dynamics and, conversely, workspace activity to coordinate distributed modules, enabling classical broadcast and ignition.

To construct the unitary reformulation of this mechanism,
we introduce a workspace-module interaction Hamiltonian
\begin{equation}
H_{WM}(t)
\;=\;
-\lambda(t)\sum_{\alpha=1}^{N_W}\sum_{a=1}^{N_M} K_{\alpha a}\, Z_\alpha Z_a.
\label{eq:H_wm}
\end{equation}
Here, $\alpha$ indexes workspace degrees of freedom
and $a$ indexes module degrees of freedom.
The coupling matrix $K_{\alpha a}$ specifies which module variables
are coordinated by the workspace, playing a role directly analogous to the classical GNW $K$ couplings.

The scalar parameter $\lambda(t)\ge 0$ controls the strength of workspace-module integration.
In a closed system, $\lambda(t)$ is an effective internal control parameter or arises from interaction with additional internal degrees of freedom via a reference structure.

Global availability is realised as entangling distributed degrees of freedom into a shared quantum state.
The term $H_{WM}$ therefore provides a quantum replacement for classical broadcast.

\paragraph{Driving term}

We introduce a non-commuting drive term of the form
\begin{equation}
H_{\mathrm{drive}}(t)
\;=\;
-\Gamma_W(t)\sum_{\alpha=1}^{N_W} X_\alpha
\;-\;
\sum_{i=1}^n \Gamma_{M_i}(t)\sum_{a\in M_i} X_a,
\label{eq:H_drive}
\end{equation}
where $X$ denotes the Pauli-$X$ operator.
The parameters $\Gamma_W(t)$ and $\Gamma_{M_i}(t)$ control the strengths of coherences
within the workspace and within each module, respectively. 

The time dependence of $\Gamma$ can be generated by embedding the control parameters into additional degrees of freedom within an extended $\HR$, of an enlarged Hamiltonian on an extended Hilbert space.

This driving term generates transitions between various $Z$-basis configurations associated with different module and workspace patterns, enabling the system to explore superpositions of basis states.

Global availability and ignition are reformulated by correlation structure introduced by the interplay between interaction terms and this non-commuting drive.

\subsection{From full quantum workspace dynamics to toy model}

The quantum Hopfield-type Hamiltonian introduced above provides an analogue of classical GNW dynamics within a fully unitary setting.

For the purposes of conceptual clarity and speculative inquiry, however, it is neither necessary nor desirable
to analyse the full many-body model in detail.

We restrict our attention to:
(i) a small number of effective degrees of freedom,
(ii) uniform or symmetric couplings
and (iii) regimes in which workspace-module interactions dominate.
In this case,the quantum Hopfield model reduces to Hamiltonians whose ground and low-energy states can be analysed explicitly.
In these limits, the dynamics naturally support globally correlated states, including GHZ-type entanglement across workspace and modules.

\section{Toy model}

\subsection{Closed unitary dynamics}
Let the entire system be described by a Hilbert space $\HU$ and a pure state $\ket{\Psi(t)}\in\HU$ evolving under a Hamiltonian $H$:
\begin{equation}
\ket{\Psi(t)} = U(t)\ket{\Psi(0)},\qquad
U(t)=e^{-iHt},\qquad
H=H^\dagger.
\end{equation}

The Hamiltonian $H$ may be taken to be the quantum Hopfield–type Hamiltonian introduced above.
There is no environment external to $\HU$.

\subsection{Hilbert Space}
This corresponds to restricting an operator algebra $\mathcal{A}$
to operators supported on $\HW \otimes \HM \otimes \HR$, where $\HW$ is the workspace Hilbert space, $\HM$ is the modules Hilbert space and $\HR$ is the reference register Hilbert space.

Note, the reference register $\HR$ allows phase-dependent transformations
to be defined and enacted.
Correlations involving $\HR$ enable phase-sensitive unitary operations
to be implemented within the closed system.

\section{Global availability as entanglement}

In classical GNW, global availability is often idealised as broadcast.
Selected information is copied into many subsystems.
In a quantum setting, however, universal copying of unknown states is forbidden by the no-cloning theorem \cite{wootters_zurek_1982}.

\subsection{Correlation distribution as an isometric embedding}
We assume a workspace state has been selected. How does it's information become globally available across the network?
Let the modules be initialised in a designated reference state
\begin{equation}
\ket{0}_M=\ket{0}_{M_1}\otimes\cdots\otimes\ket{0}_{M_n}.
\end{equation}
Let $U_{\mathrm{avail}}$ be a unitary acting on $W\otimes M_1\cdots M_n$.
This induces an isometry
\begin{equation}
V:\HW\to \HW\otimes\Big(\bigotimes_i \HMi\Big),
\qquad
V\ket{\psi}_W
=
U_{\mathrm{avail}}\big(\ket{\psi}_W\otimes\ket{0}_M\big),
\end{equation}
with $V^\dagger V=\Id_W$.

$U_{\mathrm{avail}}$ is chosen so that information initially localised in $W$ becomes distributed across  $\HW \otimes \HM $.

\subsection{Example: GHZ-type correlation distribution}\label{sec:ghzphase}

While the full quantum Hopfield workspace is a multi-qubit register with rich internal structure, we restrict attention here to a single effective workspace qubit in order to isolate the core mechanism of global availability. 

For a single qubit workspace $W$ and qubit modules $M_i$,
define controlled-NOT unitaries
\begin{equation}
U_i
=
\ket{0}\!\bra{0}_W\otimes \Id_{M_i}
+
\ket{1}\!\bra{1}_W\otimes X_{M_i},
\qquad
U_{\mathrm{avail}}=\prod_{i=1}^n U_i,
\end{equation}
as a limiting case of the more general quantum Hopfield–type architecture introduced above.

By restricting attention to a small number of degrees of freedom, assuming symmetric couplings and working in regimes where workspace–module interactions dominate over intra-module structure, the many-body Hamiltonian reduces to effective two-level dynamics.
In these limits, the ground state and low-energy states support globally correlated states, including GHZ-type entanglement across workspace and modules.

For $\ket{\psi}_W=\alpha\ket{0}+\beta\ket{1}$,
\begin{equation}
U_{\mathrm{avail}}\big(\ket{\psi}_W\ket{0\cdots 0}_M\big)
=
\alpha\ket{0}\ket{0\cdots0}
+
\beta\ket{1}\ket{1\cdots1}.
\end{equation}

In a GHZ-state, relative phase $\phi$ is absent from all single-subsystem reduced states,
yet present in operations acting on the joint systems. For the equal-amplitude case
\begin{equation}
\ket{\Psi(\phi)}
=
\frac{1}{\sqrt2}\Big(\ket{0}\ket{0^n}+e^{i\phi}\ket{1}\ket{1^n}\Big),
\end{equation}
all local marginals are maximally mixed,
\begin{equation}
\rho_W=\Tr_M\big(\ket{\Psi(\phi)}\!\bra{\Psi(\phi)}\big)=\frac{\Id}{2},
\end{equation}
while the phase appears in the following correlator on the joint system,
\begin{equation}
\expval{X_W\otimes X_{M_1}\otimes\cdots\otimes X_{M_n}}_{\Psi(\phi)}=\cos\phi.
\end{equation}

This provides a quantum analogue of global availability through internal dynamics.
Rather than a representation being copied across subsystems, entanglement exists between all subsystems (qubits).

\section{Replacing Memory: structural influence on future dynamics}

Classical memory depends on stable records.
Such records depend on decoherence and basis selection, which cannot happen in a closed quantum system.

\subsection{Memory as correlation persistence}
 Let $S$ be a source register and $A$ a memory register.
A unitary ``write'' interaction takes the form
\begin{equation}
U_{\mathrm{write}}\big(\ket{s_k}\ket{a_0}\big)=\ket{s_k}\ket{a_k},
\end{equation}
which for superposed inputs produces superpositions of correlated states $\sum_k \ket{s_k}\ket{a_k}$ .

We therefore define memory as state structures that have ongoing relevance to the dynamics.
Let $\rho_{SA}(t)$ be the reduced state on $SA$.
Subsystem $A$ functions as a memory for $S$
over an interval $t$ if correlations between $S$ and $A$ occur and can influence allowed internal operations.

Quantitatively, we can look at the mutual information
\begin{equation}
\label{eq:mutualI}
I(S{:}A)
=
S(\rho_S)+S(\rho_A)-S(\rho_{SA}),
\end{equation}
which captures total correlation. Here $S(\rho) = -\Tr(\rho \log \rho)$ denotes the von Neumann entropy.

\paragraph{Example: correlation structures as memory}

Memory therefore persists as constriants on what can and cannot happen next. 
Suppose that, following an earlier interaction, the workspace and modules occupy a GHZ-type entanglement
\begin{equation}
\ket{\Psi(\phi)}=\frac{1}{\sqrt{2}}\Big(\ket{0}\ket{0^n}+e^{i\phi}\ket{1}\ket{1^n}\Big) ,
\end{equation}
which restricts which subsequent Hamiltonian terms can act nontrivially.
In particular, any Hamiltonian supported only on a single subsystem
(e.g.\ on one module $M_i$ alone or on $W$ alone)
cannot access the phase parameter $\phi$ and cannot modify the associated global coherence, because $\phi$ is absent from all single-subsystem reduced states.
By contrast, Hamiltonian terms with appropriate joint support
(for instance, collective interactions  $X_W\otimes X_{M_1}\otimes\cdots\otimes X_{M_n}$)
can generate phase-dependent evolution.
In this way, past correlations do not store records in the traditional sense. Rather, past correlations determine which internal couplings can alter future dynamics and which couplings have no effect.

\section{Reformulating ignition: transition in correlations}

In many classical implementations of GNW, ignition is modelled as convergence towards a stable attractor generated by recurrent excitations. By contrast, finite closed quantum systems evolve unitarily and therefore do not generally exhibit attractor convergence.
Ignition can therefore be reinterpreted as a reorganisation of the joint state.

\subsection{Availability as a correlation field}
Let $W$ denote the workspace and $M_i$ the modules.
A GNW-like availability score can be defined as
\begin{equation}
\mathcal{I}=\sum_{i=1}^n I(W{:}M_i),
\end{equation}
or, more globally, as $I(W{:}M_1\cdots M_n)$, where $\mathcal{I}$ was defined in Eq. (\ref{eq:mutualI}).

\subsection{Minimal Hamiltonian}

The two-qubit Hamiltonian
\begin{equation}
H(g)= -g\,X_W X_M - h\,(Z_W+Z_M),
\qquad g\ge0,\; h>0,
\end{equation}
where $g$ is the coupling and $h$ is the local bias, can be viewed as the simplest nontrivial version of the quantum Hopfield workspace architecture, retaining only a single workspace degree of freedom, a single module, where $h$ is their dominant coupling.

For the ground state, the workspace-module $\mathcal{I}$ is 
\begin{equation}
\mathcal{I}(W:M;g) = 2 \mathcal{H}_2(p(g)),
\end{equation}
where 
\begin{equation}
H_2(p) = -p \mathrm{log}p -(1-p) \mathrm{log}(1-p)
\end{equation}
and
\begin{equation}
p(g) = \frac{1}{2}\big( 1 - \frac{2h}{\sqrt{4h^2 +g^2}}\big).
\end{equation}

Although no phase transition exists for two qubits,
this model is an analytic example
that shows hows increasing coupling $g$ increases workspace-module correlation.

Larger systems can show sharper transitions.
A canonical model is the transverse-field Ising chain \cite{sachdev_qpt,pfeuty1970ising}, whose quantum critical point is characterised by a diverging correlation length and a global reorganisation of entanglement.

Therefore, in closed quantum systems, an analogue to GNW's ignition is an increase in workspace-module correlations with increasing coupling.

\section{Discussion}

This sections looks at three implications of quantum reformulation of GNW.

\subsection{Key reformulations: Availability, memory, ignition}

To summarise, in our model

\begin{itemize}
\item Availability becomes a property of the joint system, rather than of individual subsystems.

\item Memory becomes stable internal correlation structures, rather than retrievable records. 

\item Ignition becomes an increase in entanglement, rather than convergence towards a stable attractor.

\end{itemize}

We will now look at two implications of these. Namely, that (i) access is relational rather than epistemic and (iii) meaning as relational structure. We discuss these as they are of interest to the authors artistic practice. Future work could look at wider implications, such as phenomenal equivalence, and reformulations of other models of classical consciousness/conscious access.

\subsection{Quantum conscious access is relational rather than epistemic}

In closed quantum systems, the notion of conscious \emph{access} was distinguished from outcome selection.
As we demonstrated in section \ref{sec:ghzphase}, a degree of freedom is accessible to the entire system
if it can influence the system’s subsequent internal dynamics.

In our toy model, the relative phase $\phi$ was inaccessible to local subsystems, yet operationally relevant globally through the form of the entanglement with joint, phase-sensitive operations.

Relative phase can shape future evolution through interference effects of entangled states.
In this sense, phase  is not ``known'' globally. Relative phase is operationally relevant through noncommuting joint operations that reveal interferences.

Unlike conscious access in classical GNW, access here refers to entanglement of the global state that is available to the system through its dynamics.
Therefore in this model, quantum conscious access is relational rather than epistemic.

\subsection{Meaning as relational structure}

Meaning is not usually spoken about in physics, yet it is central to art. We therefore ask what structures could give rise to meaning in our model.

Classically, stable internal states that encode information about the world form the basis of human experience. These  can generally be reported or woven into narratives. Accordingly, meaning can then be associated with representations in various ways.

In our quantum model, the physical basis for meaning must change. 
Since it cannot lie in records of experience, where, if anywhere, may it lie instead?
This is an important question because experiences that become conscious are often those that are behaviourally or personally significant, for example through memory, emotion, goals or bodily states.

Within this framework, the physical support for meaning must be the relational structure that constrains future evolution, since if meaning is accessible in our reformulated sense, it cannot reside elsewhere.
For instance, with state geometry and phases, since these determine which counterfactual evolutions are permitted or excluded.
Therefore, meaning cannot be related to represention.
Rather, it must be connected to constraints on coherent evolution, encoded in parameters that guide what the system can and cannot do next.

\section{Outlook: speculations and future directions }

\subsection{Speculations}
Let us return to the provocation that motivated this paper: what is it like to be a quantum computer? Strictly speaking, this paper's framework cannot answer this. Nevertheless, the author would still like to speculate,  both because the question motivated this work and because such speculations inform her artistic practice.

In his seminal paper {\it ``What is it like to be a bat?''} \cite{nagel1974}, Nagel wrote ``Our own experience provides the basic material for our imagination, whose range is therefore limited'', arguing that our imagination is constrained by our own forms of experience.  We tend to imagine what it is like for \emph{us} to be that other conscious entity - to possess their characteristics - rather than what it is like to be the entity itself.  

While, specific subjectivities of any individual point of view are varying degrees of unconceivable (for instance, a person with sight can never truly conceive of the experience of a person blind from birth), we have a particularly daunting task in the case of closed quantum systems. As far as we know, our brains and mental processes are best described by classical physics. As such, human experience appears to be organised around known and knowable stable states.
If consciousness could arise within a fully unitary system, as outlined above, its experiential organisation would be radically different from our own. Humans may lack the concepts needed to imagine a genuinely quantum consciousness. Any attempt to do so inevitably translates quantum processes back into familiar human forms, which limits us from fully understanding or grappling with a truly quantum point of view.

However, even if we cannot comprehend what it is like to be a quantum computer, in the same way that we can recognise the importance of echolocation to a bat's experience without knowing what echolocation feels like from within, we can still speculate about the experiential consequences of the present framework of quantum conscious access and the kinds of perceptions or modes of being it might give rise to.

A quantum-only subject would never be aware  of information in a classical sense. Conscious information in the classical GNW model assumes copying between local modules and global workspace nodes. However, in closed quantum systems the no-cloning theorem forbids copying of unknown states. A quantum-only subject would therefore never have stable records as this would require an open systems approach. Likewise, a quantum-only subject would not have any retreivable records of past events.  The past would only exist through the features of the state that continue to shape what can and cannot happen next. 

Likewise, conscious access would differ from its classical counterpart. After ignition, which has been reformulated as a widespread increase in quantum correlations across all parts, global invariants such as the relative phase would only be operationally relevant when the collective dynamics of the system were capable of revealing it through appropriate joint operations.
From this perspective, knowledge would not take the form of a collection of representations about the world. Instead, it would be embodied in the evolving structure. This suggests that any meaning or signifcance would have to arise from entanglement structure and the set of transformations that remained possible.

The author is motivated by these speculations because they stimulate imagination, creativity and open up further questions. This inquiry broadens our understanding of other possible subjectivities and the limitations of our own.  Yet investigations like this one enable us to dream, experience awe and also to emphasise with other ways of being, however unusual.

More broadly, such speculations invite us to consider that human-like forms of consciousness may be only one possible type of experience among many. Thinking about radically other-than-human forms of consciousness invites us to decentre human experience, to understand that human points of view may be just one among many  other potential forms of awareness in the universe. Even if we can never truly imagine what its like to be a bat, a quantum computer or indeed another person, thinking in these terms may remind us that our perspective is neither universal nor complete. In this sense, speculative inquiry into quantum-only forms of consciousness may ultimately tell us as much about the limitations of human understanding as it does about any possible quantum subject.

\subsection{Future Directions}

While technical, these results contribute to the growing body of artistic practice as a robust methodology for theoretical inquiry considering other-than-human intelligences \cite{2012Borgdorff, serp:morethanhuman}. For instance, ideas around distributed agency and post-anthropocentric perspectives, have all gained urgency in the context of the climate crisis and generative artificial intelligence.

This paper may inspire other artists and cultural practitioners to engage with quantum-only ontologies, where classicality never enters. This of course does not preclude representational artworks inspired by quantum theory, or from artists taking hybrid quantum-classical approaches, but it may distinguish them from works that attempt to operate within a never-classical regime.
Artworks engaging such an ontology may therefore explore quantum as configurations of quantum relations and processes.  Meaning may arise dynamically and structurally rather than symbolically.

Quantum-only ontologies therefore isolate a limiting case.
In this sense, never-classical (quantum-only), hybrid quantum–classical,
and fully classical practices can be understood as engaging different (overlapping) ontological regimes.
The framework developed here opens up one extreme of this landscape and in doing so may help inspire different processes and concepts if artists or theorists move between them.

Furthermore, this work argues for a mode of collaboration between artists and scientists that jointly explores counterfactual regimes formally.
Speculative models function less as descriptions of the world as it is and more as tools for thinking about what alternative forms of structures or experience are feasible in principle.

Finally, the framework developed here also offers a way of reflecting back on artistic practices engaging directly with quantum information systems.
In the author’s own work, long-term engagement with quantum computing has repeatedly encountered the limits of representation.
In works such as \emph{Ent-} (2022) \cite{las_ent}, recognisable forms that were animated by entanglement data do not move in a way that continues their figurative quality.
They instead move through different configurations driven by patterns of relation.
Seen through the present analysis, such practices can be understood not as failed representations nor seemingly meaningless forays into decorative abstraction, but as engagements with relational invariants such as relative phase and meaning through dynamics and structure.
However, it is important to note, the theoretical framework here does not explain the authors artworks.
Likewise, the artworks also do not illustrate the theory.
Rather, each builds upon a shared ontology and provide a language for understanding what such practices may be probing when classicality is no longer available.

\section*{Acknowledgements}

This research was supported by the British Council. I am grateful to John Goold and Hannah Andrews for insightful informal discussions and comments on various drafts of this paper.

\appendix

\section{Appendix}\label{app}
This appendix looks at the context of this current research. Firstly in section (\ref{sec:qtar}), we review organisations working across art and quantum. Secondly in section (\ref{sec:theory}), we review the theoretical art-science context around this work.

\subsection{Quantum technologies and artistic research}\label{sec:qtar}

Over the past decade, quantum science has increasingly intersected with artistic research.
Major scientific institutions and technology organisations now host artist residencies, commissions or sustained programming that explicitly recognise artists as contributors to quantum culture and discourse.
Examples include Arts at CERN \cite{artsatcern_residencies,artsatcern_collide},
art-science programmes at the University of Cambridge \cite{cavendish_artscience,qamss_art}
and artist residencies in quantum-adjacent research groups at the Niels Bohr Institute \cite{nbi_strong_art_residency}.
Artist-in-residence programmes focused on quantum technologies have also emerged through international partners,
including Studio Quantum (with strands hosted in multiple cities) \cite{goethe_studio_quantum_overview, goethe_studio_quantum_main}.
Industry research laboratories have engaged artists and creative practitioners as well,
including Google Quantum AI through an ongoing artist/creative exchange programme \cite{google_quantum_ai_creative_exchange}
and companies such as Moth Quantum explicitly stake creative adoption as their mission \cite{moth_quantum_home}.

Conversely, contemporary art organisations have shown sustained interest
in quantum science as a site of cultural and philosophical inquiry.
LAS Art Foundation, Berlin has developed the \emph{Sensing Quantum} programme \cite{las_sensing_quantum_programme},
commissioned artworks which directly used quantum computation (including the author's artwork \emph{Ent- }) (2022) \cite{las_ent}
and convened a cross disciplinary \emph{Sensing Quantum Symposium} (Berlin, 25 October 2025) \cite{las_sensing_quantum_symposium} bringing together leading artists and quantum scientists
including Laure Prouvost, Rosa Barba, Prof.\ Tommaso Calarco and Prof.\ Elham Kashefi.
More broadly, public funders and cultural organisations have articulated
the value of artists in technological development.
In July 2025 the British Council published \emph{Why technology needs artists: 40 international perspectives} \cite{britishcouncil_whytechneedsartists},
arguing, amongst other things, that artists catalyse technological advancement,
integrate cultural diversity into technical systems
and propose alternative social and technological futures. That report included quantum-focused contributions
from both artistic (including the author) and industrial research perspectives.

\subsection{Two cultures: narrowing the gap between artistic and scientific inquiry}\label{sec:theory}
Despite this growing institutional engagement,
there still remains a pronounced epistemic gap between quantum science and artistic research.
Scientific inquiry is structured around formal models with progress measured through falsifiability and predictive success.
Artistic research, by contrast, frequently operates through many modes: material experimentation,
speculative or fictional constructs,  embodied or affective modes of knowing among many others.
While these approaches may intersect,they are governed by different evaluative norms and pursue different kinds of insight and significance.

In the author’s experience leading art--science collaborations
(e.g.\ through pedagogical roles in higher education \cite{va_rca}),
if care is not taken, differences in epistemic norms can pose challenges for sustained or conceptually deep collaboration.
For instance, artists are often invited to illustrate scientific ideas, while scientists may be asked to validate artistic intuitions.
More rarely are foundational questions shared.
In particular, questions about what kinds of things or experiences might exist outside familiar biological or classical frameworks
are often treated as belonging primarily to philosophy and critical theory and are therefore approached in contemporary art through established voices (e.g.\ Haraway’s cyborg and posthuman theory, or Hayles’ analysis of posthuman subjectivity
\cite{haraway1985cyborg,hayles1999how}),
rather than being developed from scratch as formal hypotheses within art-science collaborations.

Furthermore, the questions frequently addressed within philosophical or theoretical frameworks that incorporate quantum,
including feminist science studies and agential realist accounts, explicitly incorporate classicality as measurements and decohering cuts into their ontology \cite{barad2007meeting}. 

For instance, Karen Barad's account of agential realism \cite{barad2007meeting} draws on the Copenhagen Theory of quantum mechanics to explore how meaning and matter emerge together through specific material configurations. Via quantum collapse, they give us a way to think about how the world becomes intelligible through boundary making (which they link to quantum collapse) and the resulting ethical and political implications.

Barad is deeply concerned with the moment where quantum collapses and returns a stable outcome. Their theory hinges on how meaning happens at this point of collapse, and is where classicality enters. Barad blends a quantum-classical “ethico-onto-epistem-ology” (to use their language) by design, rather than asking what happens to meaning and experience if we do not come back to the classical world. 

While these approaches have been enormously generative for contemporary art, they typically assign a foundational role to classicalisation within their ontological framework through contextual observation and material-discursive boundary formation.
For Barad, boundary-making processes were necessary for thinking through matter, meaning and knowledge as inseparable.  However, this is not the same exploration as  this paper.

\bibliographystyle{unsrt}
\bibliography{references}

\end{document}